\documentclass[aps,prd,showpacs,nofootinbib,floats,floatfix,preprintnumbers,groupedaddress,twocolumn]{revtex4}

\usepackage{bm}
\usepackage{latexsym}
\usepackage{dcolumn}
\usepackage{amsmath,amsfonts,amssymb}
\usepackage{graphicx,epsfig}

\usepackage{fancyhdr}

\def\va{Virasoro algebra}
\def\eq#1{{Eq.~(\ref{#1})}}

\begin{document}
\title{Noether current from the  surface term of gravitational action, \va\ and horizon entropy}
\author{Bibhas Ranjan Majhi}
\email{bibhas@iucaa.ernet.in}
\author{T. Padmanabhan}
\email{paddy@iucaa.ernet.in}
\affiliation{IUCAA,
Post Bag 4, Ganeshkhind,\\
Pune University Campus, Pune - 411 007, India}

\date{\today}

\begin{abstract}
We describe a simple way of obtaining horizon entropy 
using the approach based on the 
 Virasoro algebra and central charge. 
We show that the \va\ defined by the Noether currents corresponding to the \textit{surface term} of gravitational action, for the diffeomorphisms which leave the horizon structure unaltered, has a central extension that directly leads to the horizon entropy. In this approach there are no ambiguities in the calculation of the central charge. We explain why this approach is physically well motivated and could provide greater insight into the nature of horizon entropy.
\end{abstract}

\maketitle


The most intriguing feature which arises when we combine the principles of quantum theory with those of general relativity is the thermodynamic properties of horizons. There is now considerable evidence to show that these features --- which were originally thought of as properties of black holes \cite{Bekenstein:1973ur,Hawking:1974rv}--- is much more general. In fact, the emergent paradigm for gravity \cite{TPreviews} takes the thermodynamic entropy and temperature  of  null surfaces, as attributed to them by local Rindler observers, as the basis for developing the emergent perspective. These investigations also suggest that  certain level of universality in the thermodynamic properties of null surfaces is to be expected and hence approaches which compute the entropy density of a null surface in a generic, universal, manner might provide us with deeper insight into the quantum structure of spacetime. This aspect was emphasized, among others, by Carlip who has developed such an approach to compute the entropy using \va. 
(This is a generalization by Carlip \cite{Carlip:1998wz,Carlip:1999cy} of the original method by Brown and Henneaux \cite{Brown:1986nw}. A complete list of references for the later development can be found in \cite{Majhi:2011ws}.)
In this approach, one first identifies a set of vector fields based on certain physical considerations, define an algebra of charges for these vector fields and relate the entropy of the horizon to the central charge of the resulting \va\ using  Cardy's formula.
A survey of literature shows that, while the results do exhibit certain level of universality, the physical interpretation of the procedure is not completely clear. We attempt to throw more light on this issue in this letter along the following lines. 

To begin with, we will introduce the \va\ for a class of physically well motivated vector fields which are related to Noether currents.  This is partially motivated by the fact that Noether currents are known to be closely related to horizon entropy  even in more general class of theories than Einstein's gravity \cite{wald}.  

Second, we will use Noether currents
related to the diffeomorphism invariance of the York-Gibbons-Hawking \cite{ygh} surface term of the gravitational action 
rather than the usual one associated with the Einstein-Hilbert action. This is motivated by the fact that we expect the entropy to be closely related to degrees of freedom around or on the relevant null surface rather than depend on the bulk geometry of the spacetime. Once again, we know that York-Gibbons-Hawking surface term is very closely related to the entropy of the horizon; in fact, this surface term evaluated in the Euclidean extension of the local Rindler frame gives precisely one-quarter for the entropy density of the null surface. Moreover, it is known \cite{holo} that in a wide class
of gravitational theories, the surface and bulk terms of the action encode the same amount of information and hence we would expect the surface term to lead to similar results.  

Lastly, we need to identify the relevant diffeomorphisms from which the algebra has to be constructed. Here we use the criterion that the diffeomorphism should leave the near horizon form of the metric invariant in some non-singular coordinate system. Roughly speaking, such a condition reduces the symmetry of the theory from the full set of diffeomorphism to a subset which respects the existence of horizon in a given coordinate system. Since the symmetry is reduced, some of the original gauge degrees of freedom (which could have been eliminated by certain  diffeomorphisms which are now disallowed) can now be thought of as being effectively upgraded to physical degrees of freedom as far as a particular class of observers are concerned. It is these observers (like the local Rindler observers or observers at rest outside a black hole etc.) who attribute observer dependent thermodynamic properties to the horizon because of the extra degrees of freedom relevant to these observers. We will say more about this at the end of the paper. 

Given these three ingredients, the calculation of central extension of \va\ is straightforward and we obtain the entropy density of the null surface without any ambiguities in a clear and transparent manner.
We will now describe the essential ingredients of the computation.

We begin with the Noether current that can be associated with the Gibbons-Hawking surface term which is given by 
\begin{eqnarray}
A_{sur} &=& \frac{1}{8\pi G}\int_{\partial\cal{V}} \sqrt{\sigma}d^3x K
\nonumber
\\
&=& \frac{1}{8\pi G}\int_{\cal{V}} \sqrt{g}d^4x\nabla_a(KN^a)~,
\label{1.34}
\end{eqnarray}
where $N^a$ is the unit normal to the boundary $\partial{\cal{V}}$ of the region ${\cal{V}}$ while $K=-\nabla_aN^a$ is the trace of the extrinsic curvature of this boundary. Since the Lagrangian is a scalar, the Noether current $J^a$ for a diffeomorphism $x^a\rightarrow x^a+\xi^a$ can be found by considering the changes of the both sides of (\ref{1.34}) as the Lie derivative and then equating them. (Since one does not usually calculate Noether current for surface term, we have described the algebraic steps  in the Appendix, for the sake of completeness.) The conserved Noether current $J^a$ is given by:
\begin{eqnarray}
J^a[\xi] = \nabla_bJ^{ab}[\xi] = \frac{1}{8\pi G}\nabla_b\Big(K\xi^aN^b - K\xi^bN^a\Big)~.
\label{1.35}
\end{eqnarray}
where $J^{ab}$ is the Noether potential.
The corresponding charge is defined as,
\begin{eqnarray}
Q[\xi] = \frac{1}{2}\int_{\partial\Sigma} \sqrt{h}d\Sigma_{ab}J^{ab}~.
\label{1.18}
\end{eqnarray}
Here, $d\Sigma_{ab} = -d^{(d-2)}x(N_aM_b - N_bM_a)$ is the surface element of the $(d-2)$-dimensional surface $\partial\Sigma$ and $h$ is the determinant of the corresponding metric. Since our present discussion will be near the horizon, we choose the unit normals $N^a$ and $M^a$ as spacelike and timelike respectively. One can easily check that the charge in \eq{1.18}, calculated on the Killing horizon of a static metric times $2\pi/\kappa$, for the corresponding Killing vector, leads to the Bekenstein-Hawking entropy, where $\kappa$ is the surface gravity of the horizon. More generally, we get the entropy (surface) density to be one-quarter for any local Rindler horizon. 

We will next introduce the relevant algebra of charges through the definition:
\begin{eqnarray}
[Q_1,Q_2]: &=& \frac{1}{2} \Big(\delta_{\xi_1}Q[\xi_2] - \delta_{\xi_2}Q[\xi_1]\Big) 
\nonumber
\\
&=&  \frac{1}{2} \int_{\partial\Sigma} \sqrt{h}d\Sigma_{ab}\Big[\xi^a_2J^b[\xi_1] - \xi^a_1J^b[\xi_2] \Big]~,
\label{1.01}
\end{eqnarray}
where $\delta_{\xi_1}Q[\xi_2] = \int_{\Sigma} d\Sigma_a \pounds_{\xi_1}\Big(\sqrt{g}J^a[\xi_2]\Big)$ with $\partial\Sigma$ is the boundary of the ($d-1$)-hypersurface $\Sigma$.
The logic behind this definition is described in detail in \cite{Majhi:2011ws} and will not be repeated here. We merely note that it has a very natural antisymmetric structure and encodes the response of the Noether current itself to diffeomorphisms. 

We can now obtain a \va\ in the usual manner once we identify appropriate vector fields $\xi^a$. For this we shall follow the philosophy outlined earlier on and choose these vector fields to be those which leave the horizon structure invariant. To do this explicitly, let us consider the form of the metric close to the null surface in the local Rindler frame around some event. The relevant part of the metric in the $x-t$ plane is given in the standard Rindler coordinates by:
\begin{eqnarray}
ds^2 = -2\kappa x dt^2 + \frac{1}{2\kappa x} dx^2 + dx_\perp^2  ~.
\label{1.03}
\end{eqnarray}
where the Killing horizon is at $x = 0$ and $x_\perp$ denotes the transverse coordinates. We will first transform to Bondi-like (Gaussian null) coordinates by the transformation
\begin{eqnarray}
du = dt - \frac{dx}{2\kappa x}; \,\,\
dX = dx~.
\label{1.07}
\end{eqnarray}
thereby transforming the metric to 
\begin{eqnarray}
ds^2 = -2\kappa X du^2 - 2 du dX+ dx_\perp^2 ~.
\label{1.08}
\end{eqnarray}
The Killing horizon in this frame is at $X=0$.
We now choose our vector fields $\xi^a$ by imposing the condition that the metric coefficients
 $g_{XX}$ and $g_{uX}$
 in the above expression remain unchanged:
\begin{eqnarray}
\pounds_{\xi} g_{XX} = 0; \,\,\,\ \pounds_{\xi} g_{uX} =0~.
\label{1.11}
\end{eqnarray}
which gives
\begin{eqnarray}
&&\pounds_{\xi} g_{XX} = -2\partial_X{\xi}^u = 0;
\nonumber
\\
&& \pounds_{\xi} g_{uX} = -\partial_u {\xi}^u - 2\kappa X \partial_X {\xi}^u - \partial_X {\xi}^X =0~.
\label{1.12}
\end{eqnarray}
The solutions to these equations are given by:
\begin{eqnarray}
{\xi}^u = F(u, x_{\perp}); \,\,\
{\xi}^X = - X\partial_u F(u,x_{\perp})~,
\label{1.13}
\end{eqnarray}
 The  condition $\pounds_{\xi}g_{uu} = 0$ is satisfied near the horizon, since a direct calculation yields $\pounds_{\xi}g_{uu} = {\cal{O}}(X)$. (These gauge conditions have appeared earlier in \cite{Tanabe:2011zt} in the context of late time symmetry near the horizon of a black hole.)
In the original Rindler  coordinates these vector fields  reduce to the following form:
\begin{eqnarray}
\tilde{\xi}^t = T - \frac{1}{2\kappa}\partial_tT; \,\,\,\ \tilde{\xi}^x = -x\partial_t T~,
\label{1.17}
\end{eqnarray}
where $T(t,x,x_{\perp}) \equiv F(u,x_{\perp})$.

We now have all the ingredients (diffeomorphism vector fields $\xi^a$, Noether currents and the definition of algebra of charges) to proceed with the computation. For the Rindler metric with 
\begin{eqnarray}
&&N^a = (0,\sqrt{2\kappa x}); \,\,\ M^a = (\frac{1}{\sqrt{2\kappa x}},0);
\nonumber
\\
&&K= -\sqrt{\frac{\kappa}{2x}}
\label{new1}
\end{eqnarray}
and $d=4$,
substituting \eq{1.17} in \eq{1.18} and \eq{1.01} we obtain
\begin{eqnarray}
Q[\xi] = \frac{1}{8\pi G}\int_{\cal{H}} \sqrt{h}d^2x\Big(\kappa T - \frac{1}{2}\partial_t T\Big)~,
\label{1.37}
\end{eqnarray}
Thus for a given choice of $T$ we get a corresponding vector fields $\xi^a$ (see \eq{1.17}) and charge $Q$ (see \eq{1.37}). Since both are linear in $T$, if we expand the function $T$ in terms of a set of basis functions $T_m$, with
\begin{eqnarray}
T = \displaystyle\sum_m A_m T_m;\quad A_m^* = A_{-m}
\label{1.25}
\end{eqnarray}
we will get corresponding expansions for $\xi^a$ (in terms of some $\xi_m^a$-s defined by \eq{1.17} with $T$ replaced by $T_m$) and for $Q$ (in terms of some $Q_m$-s defined by \eq{1.37} with $T$ replaced by $T_m$). The usual procedure is now to choose the basis $T_m$ such that the resulting $\xi_m^a$ obeys the algebra:
\begin{eqnarray}
i\{\xi_m,\xi_n\}^a = (m-n)\xi_{m+n}^a~,
\label{new2}
\end{eqnarray}
where $\{,\}$ is the  Lie bracket. This can be achieved by the choice
\begin{eqnarray}
T_m = \frac{1}{\alpha}\exp\left[im\left(\alpha t  + g(x) + p.x_{\perp}\right)\right]~,
\label{1.28}
\end{eqnarray}
where $,\alpha$ is a constant, $p$ is an integer and $g(x) = G(X=x)= - \alpha\int\frac{dx}{2\kappa x}$. (Such a choice is standard in these computations and has been used several times in the literature; see, for example,   \cite{Silva:2002jq}.) 
Since transverse directions are non-compact due to our Rindler approximations, we will assume that 
$T_m$ is periodic in the transverse coordinates, with  say the periodicities  $L_y$ and $L_z$ on $y$ and $z$ respectively.
It is now straightforward to compute the algebra of the charges corresponding to $T=T_m,T_n$, say, by using \eq{1.01} and we get:
\begin{eqnarray}
[Q_m,Q_n]: &&= \frac{1}{8\pi G}\int_{\cal{H}} \sqrt{h}d^2x \Big[\kappa(T_m\partial_tT_n - T_n\partial_t T_m) 
\nonumber
\\
&&- \frac{1}{2}(T_m\partial_t^2T_n - T_n\partial_t^2 T_m)
\nonumber
\\
&&+\frac{1}{4\kappa}(\partial_tT_m\partial_t^2T_n - \partial_tT_n\partial_t^2 T_m)\Big]~.
\label{1.40}
\end{eqnarray}
We now substitute \eq{1.28} in \eq{1.37} (with $T$ replaced by $T_m$) and in \eq{1.40} and integrate over a cross-section area $A=L_yL_z$ in transverse directions to obtain the explicit expressions:

\begin{eqnarray}
&&Q_m = \frac{A}{8\pi G}\frac{\kappa}{\alpha}\delta_{m,0};
\nonumber
\\
&&[Q_m,Q_n]=  -\frac{i\kappa A}{8\pi G\alpha}(m-n)\delta_{m+n,0}
\nonumber
\\
&&- im^3 \frac{\alpha A}{16\pi G\kappa}\delta_{m+n,0}~. 
\label{1.29}
\end{eqnarray}
Therefore, the central term in the algebra  is:
\begin{eqnarray}
K[\xi_m,\xi_n] &=& [Q_m,Q_n] + i(m-n) Q_{m+n} 
\nonumber
\\
&=&  -im^3 \frac{A}{16\pi G}\frac{\alpha}{\kappa}\delta_{m+n,0}~.
\label{1.41}
\end{eqnarray}
from which we can read off the central charge $C$ and the zero mode energy $Q_0$  as:
\begin{eqnarray}
\frac{C}{12} =  \frac{A}{16\pi G}\frac{\alpha}{\kappa}; \,\,\,\ Q_0 = \frac{A}{8\pi G}\frac{\kappa}{\alpha}
\label{1.42}
\end{eqnarray}
Finally, we use the Cardy formula \cite{Cardy:1986ie,Carlip:1998qw} to obtain the entropy:
\begin{eqnarray}
S = 2\pi\sqrt{\frac{CQ_0}{6}} = \frac{A}{4G}~.
\label{1.43}
\end{eqnarray}
This is exactly the result we would have expected.

Before discussing the broader and conceptual aspects of this result we 
would like to briefly mention one technical point. The idea of choosing 
the diffeomorphisms which preserve the horizon structure can be implemented in 
many different ways. Our motivation was to choose a minimal set of vector fields $\xi^a$ which 
will allow us to generate the appropriate Virasoro algebra. It is possible to make other 
choices in such a way that the vectors $\xi^a$ has appropriate limiting behaviour close to the horizon (like e.g suitable powers of the distance from the horizon etc.) and implement the same idea. All such
diffeomorphisms will lead to the same result. (Of course if the metric has 
to be left completely invariant then the diffeomorphism has to be 
generated by a Killing vector which will not exist in a general spacetime.) 
The basic idea is  to impose a \textit{minimal} set of conditions on the 
vector fields generating the diffeomorphism, which will lead to the necessary  Virasoro algebra. This is 
precisely what we have achieved.

The logical simplicity of the above approach is especially noteworthy. Just to provide a contrast we will mention couple of technical points related to previous derivations in the literature leading to similar results. In most of the earlier work (for a complete list, see \cite{Majhi:2011ws}), 
to obtain the correct entropy  one had to either shift the zero mode energy (as done in. e.g., \cite{Carlip:1999cy}) or choose a parameter contained in the Fourier modes of $T$ (for instance $\alpha$ in \eq{1.28} here) as the surface gravity (as done in, e.g.,\cite{Silva:2002jq}) or both \cite{Majhi:2011ws} depending on the action of the theory we have started with. \textit{Here, interestingly, we did not require any such ad hoc prescription}. This is because the parameter $\alpha$ in the expression of $T_m$ \eq{1.28} did \textit{not} appear in the final expression of entropy. Hence the present derivation of entropy is a lot simpler compared to others and free of any ambiguity. 

As far as we know, using (a) the Noether current corresponding to the York-Gibbons-Hawking surface term and (b) choosing the vector fields by demanding invariance of the Rindler form of the metric, have not been attempted before and are new features of this work. This is important because it further strengthens the peculiar feature present in all gravitational actions \cite{holo}, viz., that the same information is encoded in both bulk and boundary terms of the action. Our analysis can also be considered as local in the sense that it uses features related to the null surface both in the choice of surface term in the action and in the choice of specific diffeomorphisms. It also provides a more transparent physical interpretation of some of the mathematical steps involved in this approach. 

Let us briefly summarize what has been achieved in this letter. We consider a class of diffeomorphisms which leave a particular form of the Rindler metric invariant. We also introduce the Noether current associated with the surface term in the action functional for gravity (in contrast to previous approaches in the literature in which either the bulk term or the  Einstein-Hilbert action was used to define the Noether current). Given the Noether current and diffeomorphism,  there is a natural \va\ which can be associated with the horizon. We compute the central extension of this \va\ and show that it leads --- via Cardy formula --- to the correct entropy density of the horizon. The class of diffeomorphisms which keep the horizon structure invariant is a subset of all possible diffeomorphisms without any constraint. From a physical point of view it seems natural to assume that such a constraint upgrades the remaining degrees of freedom (which were originally pure gauge degrees of freedom) to effectively real degrees of freedom for the observer who perceives the horizon. This naturally accounts for an \textit{observer dependent} notion of degrees of freedom which contributes to an observer dependent entropy.

The last point is conceptually important and provides a nice, simple, physical picture of the connection between horizon entropy and the degrees of freedom which contribute to it. In conventional physics, we are accustomed to thinking of degrees of freedom of a system as absolute and independent of observer or the coordinate system used by the observer. In such a picture, the entropy --- which is related to the logarithm of the degrees of freedom --- will also be absolute and independent of the observer. We however know that horizon entropy, horizon temperature etc. must be treated as observer dependent notions \cite{comment1} because a freely falling observer through a black hole horizon and a static observer outside the black hole will attribute different thermodynamic properties to the horizon. \textit{It follows that any microscopic degrees of freedom which leads to horizon entropy must also be necessarily  observer and coordinate dependent.}
So the question ``what are the degrees of freedom responsible for black hole entropy?'' has no observer independent answer. We find that the notion of residual gauge symmetries which are perceived as real degrees of freedom under a restricted class of diffeomorphism, allows us to quantify this notion. An observer who perceives a horizon works with a theory having less diffeomorphism symmetry if she wants to retain the structure of the metric near the horizon. This necessarily upgrades some of the original gauge degrees of freedom to effectively true degrees of freedom as far as this particular observer is concerned. As a result, this particular class of observers will attribute an entropy to the horizon in an observer dependent manner.

\section*{Acknowledgement}

The research of TP is partially supported by the JC Bose research grant of DST India.

\appendix
\section{\label{AppendixA}Derivation of noether current}
\renewcommand{\theequation}{A.\arabic{equation}}
\setcounter{equation}{0}  
  Consider a general Lagrangian which is a total derivative of a vector field so that the resulting action has only a surface contribution. Then the Lagrangian density can be expressed as
\begin{eqnarray}
\sqrt{g}L = \sqrt{g}\nabla_aA^a~,
\label{App1}
\end{eqnarray}
where $L$ is a scalar.
Under a diffeomorphism $x^a\rightarrow x^a+\xi^a$  the left hand side changes by:
\begin{eqnarray}
\delta_{\xi}\Big(\sqrt{g}L\Big) \equiv \pounds_{\xi} \Big(\sqrt{g}L\Big) = \sqrt{g}\nabla_a\Big(L\xi^a\Big)~.
\label{App2}
\end{eqnarray}
On the other hand, the variation of the right hand side of (\ref{App1}) is given by:
\begin{eqnarray}
&&\delta_{\xi}\Big(\sqrt{g}\nabla_aA^a\Big) = \pounds_{\xi}\Big[\partial_a\Big(\sqrt{g}A^a\Big)\Big]
\nonumber
\\
&& = \partial_a\Big[A^a\pounds_\xi\sqrt{g} + \sqrt{g}\pounds_\xi A^a\Big]
\nonumber
\\
&& = \sqrt{g}\nabla_a\Big[\nabla_b\Big(A^a\xi^b\Big) - A^b\nabla_b\xi^a\Big]
\label{App3}
\end{eqnarray}
 Equating (\ref{App2}) and (\ref{App3}) we obtain the conservation law $\nabla_a J^a =0$ for the Noether current:
\begin{eqnarray}
J^a[\xi] = L\xi^a - \nabla_b(A^a\xi^b) + A^b\nabla_b\xi^a~,
\label{App5}
\end{eqnarray}
which after using (\ref{App1}) reduces to the following form:
\begin{eqnarray}
J^a[\xi] = \nabla_bJ^{ab}[\xi] = \nabla_b\Big[\xi^aA^b-\xi^bA^a\Big]~.
\label{App6}
\end{eqnarray}
This was the result used in the paper.




\begin{thebibliography}{99}
\bibitem{comment1} In fact the Davis-Unruh effect suggests that all thermodynamic variables (like the temperature, entropy etc. of a glass of water) must be treated as observer dependent. This follows from the fact that a glass of water is a highly excited state of the vacuum and if inertial and Rindler observers assign different temperatures, say, to the vacuum they will also assign different temperatures to the glass of water.
\end{thebibliography}
\end{document}